\documentstyle[]{l-aa}
\input{psfig}
\begin{document}
  \thesaurus{06(06.09.1;
              06.15.1;02.18.8)}
   \title{Relativistic effects in the solar EOS}
   \subtitle{A helioseismic forward analysis}
   \author{A. Bonanno\inst{1}\and A.L. Murabito\inst{2}\and\, L. Patern\`o\inst{2}}
\offprints{A. Bonanno}
\institute{Osservatorio Astrofisico di Catania, Citt\`a Universitaria, I-95123 Catania, Italy\\
           (abo@sunct.ct.astro.it)
	        \and
            Dipartimento di Fisica e Astronomia dell'Universit\`a, Sezione Astrofisica, Citt\`a Universitaria, I-95123 Catania, Italy\\ 
(amu@sunct.ct.astro.it, lpaterno@alpha4.ct.astro.it)}
\date{Received 18 April 2001 / Accepted 14 June 2001}
   \maketitle

   \begin{abstract}
We study the sensitivity of the sound speed to relativistic corrections of the equation of state (EOS) 
in the standard solar model by means of a  helioseismic forward analysis.
We use the latest GOLF/SOHO data for $\ell = 0,1,2,3$ modes to confirm that 
the inclusion of the relativistic corrections to the adiabatic exponent $\Gamma_1$ computed 
from both OPAL and MHD EOS leads to a more reliable theoretical 
modelling of the innermost layers of the Sun. 
      \keywords{Sun: interior --
                Sun:  oscillations --
                Equation of state}
   \end{abstract}

\section{Introduction}
It has recently been shown (Elliott \& Kosovichev \cite{Elliot}) 
that the inclusion of relativistic effects in 
the equation of state (EOS) leads to a very good agreement between the solar models and 
the seismic Sun. In particular, the inversions of SOI-MDI/SOHO $p$-mode frequencies 
for the adiabatic exponent
$\Gamma_1$ show that MHD EOS reproduces the interior of the Sun with 
great accuracy, when the relativistic contribution to the Fermi-Dirac statistics is included.

It is thus interesting to approach the same problem by means of the forward analysis by comparing
the theoretical eigenfrequencies with the observed ones. 
Unfortunately this method is not directly 
applicable since our description of the outer layers of the Sun is still far from 
complete and many theoretical uncertainties would influence our conclusions.
However, since such small effects in solar EOS are most important only in the deep interior, it is
possible to make use of the acustic mode frequency small separation diagnostic,  
$\delta\nu_{\mathrm{\ell,n}}=\nu_{\mathrm{\ell,n}}-\nu_{\mathrm{\ell+2,n-1}}$, 
for spherical harmonic degrees $\ell = 0,1$ and radial order ${\rm n}\gg\ell$ (Tassoul \cite{Tassoul}). 
The main property of this quantity is that it is strongly sensitive to the sound speed 
gradient near the solar centre while it is weakly dependent on the details of the 
treatment of the outer layers. Since the relativistic effects manifest themselves mainly through a
depletion of $0.1\% - 0.2\%$ of the adiabatic index $\Gamma_1$, 
we expect a quantitatively similar change of sound speed gradient in the solar core.

The acoustic mode frequency small separation analysis has recently been used for 
estimating the seismic age of the Sun (Dziembowski {\it et al.} \cite{Dziem}) and the related 
implications of the uncertainties in the $\rm S_{11}$ astrophysical factor determinations
(Bonanno \& Patern\`o \cite{Bonanno}).

Here we show that the mentioned above analysis can also be used to verify how 
the different physical characteristics of the MHD and OPAL EOS reflect on the accuracy of the description of the
stratification of the internal layers of the Sun. 
On performing a $\chi^2$ analysis  of the 
latest published GOLF/SOHO data for different solar models, 
we confirm the main conclusion of Elliott \& Kosovichev (\cite{Elliot}), based on an inversion
analysis, that the inclusion
of the relativistic effects in the EOS is in any case required to improve the accuracy of solar
models, independent of which EOS is used.

\section{The solar model}
In our analysis we used the GARching SOlar Model 
(GARSOM) code which has been described in detail in Schlattl {\it et al.} (\cite{Schlattl}). 
It includes the latest OPAL-opacities and either OPAL or MHD EOS, 
and it takes into account the microscopic diffusion of the elements heavier than hydrogen. 
Our standard solar model has been verified in detail 
in Turck-Chi\`eze {\it et al.} (\cite{gong98}) and found
in good agreement with other up-to-date solar models, and,
in particular, it is consistent with the observed $\rm L_\odot$ 
and $\rm T_{eff}$ within $10^{-4}$, at an age of $\rm 4.60\,Gy$, adopting the surface value $\rm Z/X=0.0245$. 
We then included the relativistic correction leading term to the 
adiabatic index $\Gamma_1$ derived from the relativistic 
evaluation of the Fermi-Dirac integrals of the EOS
in the solar core by means of the expression  (Elliott \& Kosovichev \cite{Elliot}):
\begin{equation}
\rm \frac{\delta\Gamma_1}{\Gamma_1}\simeq -\widetilde{T}\,\frac{2+2X}{3+5X}
\end{equation}
where $\rm \widetilde{T}$ is a dimensionless temperature in units of ${\rm m_{e}}c^2/k$, with $\rm m_{e}$ the electron mass, $c$ the light speed in vacuum, $k$ the Boltzmann constant, and $\rm X$ the hydrogen abundance by mass. 
As expected, the correction to $\Gamma_1$ is negative, namely $\Gamma_{1,\rm rel} < \Gamma_{1,\rm nr}$, since $\Gamma_1$ tends to shift from the non-relativistic value of 5/3 to the extremely relativistic one of 4/3. 
 
The corresponding relativistic corrections to the leading terms for sound speed, $c_s$, and density, $\varrho$, are respectively:
\begin{equation}
\frac{\delta c_s}{c_s} \simeq \frac{1}{2}\,\frac{\delta \Gamma_1}{\Gamma_1} - \frac{15}{64\sqrt{2}}\,\rm \widetilde{T}\rm e^{\psi}
\end{equation}
and 
\begin{equation}
\frac{\delta \varrho}{\varrho} \simeq \frac{15}{8}\,\rm \widetilde{T}\left(1+\frac{\rm e^{\psi}}{4\sqrt{2}}\right)
\end{equation}
where $\psi$ is the degeneracy parameter, that is about -1.14 at the Sun's centre, and decreases  noticeably toward the surface, the partial degeneracy being completely removed at $0.4\,\rm R_\odot$. 

The behaviour, as functions of the fractional radius, of the relative differences between the quantities $\Gamma_1$, $c_s$ and
$\varrho$ calculated with relativistic corrections and without them is shown in Fig. 1. 

The term 
$-(15/64\sqrt{2})\rm \widetilde{T}\rm e^{\psi} = \delta \rm P/\rm P - \delta \varrho/\varrho$, in Eq.(2) is negligible with respect to $\delta \Gamma_1/\Gamma_1$ indicating that the relativistic corrections to the pressure, P, and density, $\varrho$, cancel each other almost completely and the correction to $c_s$ is entirely dominated by the correction to $\Gamma_1$. Also the term
$\rm e^{\psi}/4\sqrt{2}$ in Eq.(3) is negligible with respect to unity, indicating that in the solar case the coupling between degeneracy and relativistic effects is weak.
\begin{figure}[!ht]
\centerline{\psfig{file=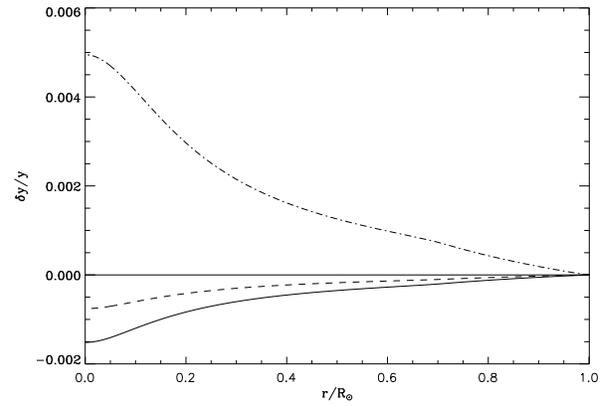,width=8cm}}
\caption{Behaviour, as functions of the fractional radius, of the relative differences between relativistic and non-relativistic quantities $\delta {\rm y/y = (y_{rel}-y_{nr})/y_{nr}}$, where the ys stand for $\Gamma_1$ (continuous line), $c_s$ (dashed line), and $\varrho$ (dashed-dotted line) respectively.}
\end{figure}

\section{Results with GOLF/SOHO data}
We used the latest GOLF/SOHO data for $\ell = 0,1,2,3$ 
 obtained with long time series and by 
taking into account the asymmetric line profile in data reduction (Thiery {\it et al.} \cite{Thiery}). 
In particular, we determined the acoustic mode small spacing 
difference $\delta\nu_{\mathrm{\ell,n}}$ for $\ell = 0,1$ and ${\rm n}\gg\ell$ for our solar model, 
and studied the difference $\delta\nu_{{i,\rm {n}},\odot}-\delta\nu_{{i,{\rm n,model}}}$ between  data and model. We then constructed the two $\chi^2$ indicators 
(Dziembowski {\it et al.} \cite{Dziem}, Schlattl {\it et al.} \cite{Schlattl})  
\begin{equation}
 \chi^2_i=\frac{1}{\rm M-m+1}\sum_{\rm {n=m}}^{\rm M}\frac{(\delta\nu_{{i,{\rm n}},
\odot}-\delta\nu_{{i,{\rm n,model}}})^2}{\sigma^2_{{i,\rm{n}}}+\sigma^2_{{2+i,\rm{n-1}}}}
\label{eq2}
\end{equation}
where $i$ stands for $\ell = 0,1$, m = 10 and M = 26. 
Fig. 2 and Fig. 3 show the behaviour of the terms in the sum  defined in
Eq.(4) in non-relativistic and relativistic cases for MHD and OPAL EOS respectively.
\begin{figure}[!hb]
\centerline{\psfig{file=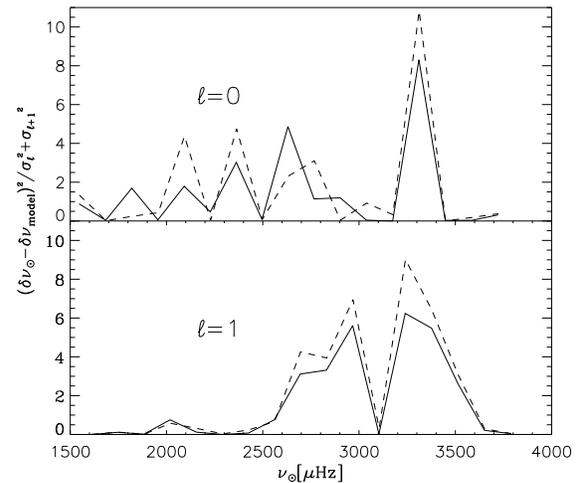,width=8cm}}
\bigskip
\caption{Relativistic  (continuous line) and non-relativistic (dashed line) contribution 
to the $\chi^2$ calculation for MHD EOS.}
\end{figure}
\begin{figure}[!ht]
\centerline{\psfig{file=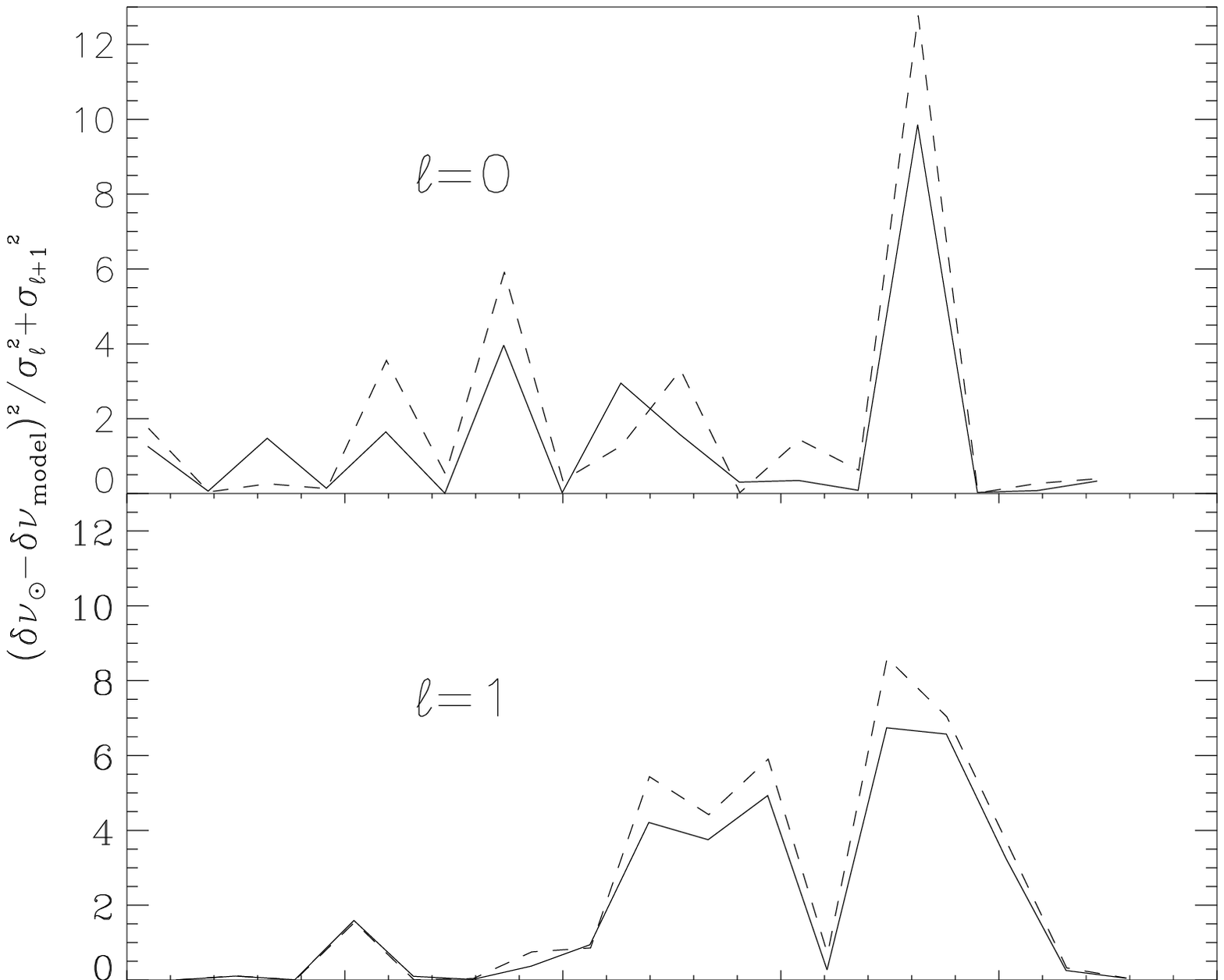,width=8cm}}
\bigskip
\caption{Relativistic  (continuous line) and non-relativistic (dashed line) contribution 
to the $\chi^2$ calculation for OPAL EOS.}
\end{figure}
\begin{table}
\caption{$\chi^2$ results in non relativistic (NR) and relativistic (REL) cases for MHD and OPAL EOS.}
\label {T1}
\center
\begin{tabular}{ccccc}
\hline 
\smallskip
EOS & $\chi^2_0$(NR) & $\chi^2_0$(REL) & $\chi^2_1$(NR) & $\chi^2_1$(REL)\\
\hline
MHD & 1.73 & 1.41 & 2.13 & 1.67\\
OPAL & 1.91 & 1.41 & 2.32 & 1.94\\
\hline
\end{tabular}
\end{table}
The difference between the relativistic and non relativistic case
is larger for $\ell=0$ in the frequency range 
beetwen $2000$ and $2500$ $\mu$Hz,
and for $\ell=1$ beetwen $2500$ and $3000$ $\mu$Hz.
The $\chi^2$ results are shown in Table \ref{T1} 
where it is possible to note that the models with relativistic corrections have rather smaller 
$\chi^2\,$s and there is no significant difference
between $\chi^2_0$ and $\chi^2_1$ calculated for OPAL
and MHD EOS. However, MHD EOS appears to be slightly favoured with respect to OPAL EOS.

\section{Conclusions}
Our results show that the  acoustic mode  frequency small separations are 
sensitive to the inclusion of the relativistic 
effects. It would be interesting to discuss the relevance of these effects in
the helioseismic determination of the solar age
and related problems with $\rm S_{11}$ uncertainties. We plan to 
address this issue in a forthcoming communication.

\begin{acknowledgements}
\item We are most grateful to H. Schlattl for useful discussions during the preparation of the manuscript.
\end{acknowledgements}

\end{document}